\documentclass[superscriptaddress,nofootinbib,prd,11pt]{revtex4}
\usepackage{graphicx}
\usepackage{cancel}
\usepackage{amssymb}
\usepackage{textcomp}
\usepackage{amsmath}
\usepackage{bm}
\usepackage{times}
\usepackage{color}

\begin{document}

\title{  Anti-de Sitter massless scalar field spacetimes in arbitrary dimensions}
\author{Sebasti\'an Garc\'{\i}a S\'aenz}
\affiliation{Department of Physics, Columbia University, New York, NY 10027, USA}
\email{sg2947@columbia.edu}
\author{Cristi\'an Mart\'{\i}nez}
\affiliation{
 Centro de Estudios Cient\'{\i}ficos (CECs), Av.\ Arturo Prat 514, Valdivia, Chile \\ and 
Universidad Andr\'es Bello, Av.\ Rep\'{u}blica 440, Santiago, Chile}
\email{martinez@cecs.cl}

\begin{abstract}
We consider $d$-dimensional static spacetimes in Einstein gravity with a cosmological constant in the presence of a minimally coupled massless scalar field. The spacetimes have a $(d-2)$-dimensional base manifold given by an Einstein space and the massless scalar field depends only on the radial coordinate. The field equations are decoupled in the general case, and can be solved exactly for the cases when either the cosmological constant vanishes or the base manifold is Ricci flat. We focus on the case of a negative cosmological constant and a Ricci-flat base manifold. The solution has a curvature singularity located at the origin, where also the scalar field diverges. Since there is no event horizon surrounding this singularity, the solution describes a naked singularity dressed with a nontrivial scalar field. This spacetime is an asymptotically locally anti-de Sitter one when the Ricci-flat base manifold is locally flat. The asymptotic solution for an arbitrary Einstein base manifold is found and the corresponding mass, calculated through the canonical generator of the time-translation invariance, is shown to be finite. The contribution to the mass from the scalar field at infinity is also discussed. 
\end{abstract}
\maketitle
\section{Introduction}

Although no elementary scalar field has been discovered to date, such fields are predicted to exist in a number of different theories. In spite of this, in general relativity, the so-called ``no-hair'' theorem rules out the existence of black holes in the presence of a minimally coupled scalar field, suggesting a possible incompatibility between gravity and scalar fields at least for asymptotically flat spacetimes. However, the AdS/CFT correspondence and its recent developments, e.g.\ holographic superconductors, have motived the search of solutions dressed with a scalar field that asymptotically approach the anti-de Sitter (AdS) spacetime. In the presence of a negative cosmological constant the situation is completely different: AdS spacetimes are stable against scalar field perturbations even in the case of self-interaction potentials unbounded from below, provided that the mass term fulfills the Breitenlohner-Freedman bound \cite{BF,MT}. In such a setup, which is not possible in asymptotically flat spacetimes, the standard no-hair theorem does not hold. Another way for circumventing this theorem is to consider a nonminimal conformal coupling and a negative cosmological constant. In this case, a mass term, which is proportional to a negative constant Ricci scalar, appears in the scalar field equation. This was implemented in three dimensions\footnote{In four dimensions, without a cosmological constant, the addition of the conformal coupling $\frac{1}{6}R \phi^2$ term to the action yields a spherically symmetric solution, in which the scalar field diverges at the event horizon \cite{BBMB}.} yielding an exact black hole solution dressed with a regular scalar field \cite{Martinez:1996gn}. In the minimal coupling case with a self-interaction potential, exact hairy black holes have been found in three \cite{Henneaux:2002wm,Correa:2011dt} and four spacetime dimensions \cite{Martinez:2004nb}. In these solutions the cosmological constant plays a key role since, in particular, a negative cosmological constant opens up the possibility of having black holes in vacuum with nonspherical event horizons \cite{Lemos:1994xp,Vanzo:1997gw,Brill:1997mf}. Some exact scalar hairy black hole solutions in diverse spacetime dimensions, in this class of models, can be found in \cite{others}. See \cite{Hosler:2009sf} for a comprehensive review and additional references.

In this article we study $d$-dimensional static solutions of the Einstein field equations in the presence of a minimally coupled massless scalar field and a negative cosmological constant. The absence of a self-interaction potential or a nonminimal coupling restricts the possibility of finding black hole solutions. However, the problem deserves proper attention since it represents the most simple way of coupling matter to gravity with a negative cosmological constant. Moreover, as it has been discussed in the literature (see e.g.\ \cite{Gubser:2000nd}, \cite{Das:2001rk} and \cite{Husain:2002jr}), asymptotically AdS spacetimes containing naked singularities could play a role in the context of the AdS/CFT correspondence. 

For the case of a vanishing cosmological constant, the general static and spherically symmetric solution of this model was first found in four dimensions by Fisher \cite{Fisher:1948yn}, and later rediscovered in \cite{Janis:1968zz} and \cite{Wyman:1981bd} (see also \cite{Virbhadra:1997ie}). The instability of the Fisher solution under spherical perturbations was recently proven in \cite{Bronnikov:2011if}. The higher-dimensional generalization was found by Xanthopoulos and Zannias in \cite{Xanthopoulos:1989kb}, and further studied in detail in \cite{Abdolrahimi:2009dc}, while the solution for the three-dimensional case was reported in \cite{Barrow1986}, \cite{Virbhadra:1994xz}, and \cite{Clement:1998an}. Regardless of the number of spacetime dimensions, the results dictate that the static and spherically symmetric solution, with a nontrivial massless scalar field, corresponds to a spacetime containing a naked singularity, as it is expected by virtue of the no-hair theorem for asymptotically flat spacetimes.

The inclusion of a negative cosmological constant was first considered in \cite{Clement:1999jz}, where the general solution for $d=3$ spacetime dimensions was found, and later independently rediscovered in \cite{Das:2001rk} and \cite{Daghan:2005yn}. In four dimensions we can mention the existence of a particular plane-symmetric solution with nonzero cosmological constant (of arbitrary sign) given in \cite{Vuille:2007ws}. Finally, a particular solution to the problem in $d$ dimensions with a flat base manifold was found in \cite{Husain:2002jr}. As far as we know, there are no general exact results for this model in higher dimensions. The main purpose of this article is to generalize previous results in two different ways: (i) by including a nonvanishing cosmological constant term in arbitrary dimension, and (ii) by studying the case in which the base manifold, i.e.\ the boundary of the spacelike sections, is a $(d-2)$-dimensional Einstein manifold rather than the usual $(d-2)$-sphere. An additional goal in this work is to determine the mass of the configurations and analyze the contribution of the asymptotic value of the scalar field on the mass.

The paper is organized as follows. In Sec.\ \ref{sec:field equations}, after deriving the field equations, we introduce an appropriate variable that allows the field equations to become a decoupled system of differential equations. Next, we classify the cases in which an exact integration of the equations can be done. One case corresponds to a solution where the cosmological constant is negative and the base manifold (an Einstein space) is Ricci flat. Section \ref{exactsolution} is devoted to the study of this particular case and the corresponding general exact solution is found and analyzed. For the general case, where the cosmological constant is nonzero and the base manifold is not a Ricci-flat one, an exact solution is not available. However, as we show in Sec.\ \ref{asolution}, the asymptotic solution can be found regardless of the value of the curvature of the Einstein base manifold. Using the expression for the canonical generator associated with the time-translation invariance of the system, we present in Sec.\ \ref{sec:mass} the computation of the mass of the solutions having the asymptotic behavior determined in the previous section. In particular, the mass of the exact solution discussed in Sec.\ \ref{exactsolution} is given. We also discuss the contribution to the mass coming from a nonvanishing value of the scalar field at infinity. Finally, some general remarks are given in Sec.\ \ref{sec:remarks}.

\section{Action and field equations}  \label{sec:field equations}

We consider a real massless scalar field minimally coupled to Einstein gravity in $d>2$ spacetime dimensions in the presence of a cosmological constant $\Lambda$. The action for this model is given by
\begin{equation} \label{eq:action}
I[g_{\mu\nu},\phi]=\int d^dx\sqrt{-g}\left(\frac{R-2\Lambda}{2\kappa}-\frac{1}{2}g^{\mu\nu}\partial_{\mu}\phi\partial_{\nu}\phi\right),
\end{equation}
where $\kappa$ is the Einstein constant. The corresponding field equations are
\begin{equation}
R^{\mu}_{~~\nu}-\frac{2\Lambda}{d-2}\delta^{\mu}_{~~\nu}=\kappa\partial^{\mu}\phi\partial_{\nu}\phi,
\end{equation}
and
\begin{equation} \label{KG}
\Box \phi=0.
\end{equation}
We are interested in static configurations defined by the following Ansatz:
\begin{equation} \label{eq:metric}
ds^2=-e^{2h(r)}f^2(r)dt^2+\frac{dr^2}{f^2(r)}+r^2\gamma_{mn}dz^mdz^n, \quad \mbox{with} \quad \phi=\phi(r).
\end{equation}
Here $\gamma_{mn}$ is the metric of a $(d-2)$-dimensional Einstein manifold $\Sigma$ of Euclidean signature, whose Ricci tensor is given by $R_{\Sigma~~n}^{~~m}=(d-3)\gamma\delta^m_{~~n}$. The constant $\gamma$ can be taken to be either $0$, $+1$ or $-1$. The manifold $\Sigma$ is assumed to be nonsingular and to have a finite volume, denoted by $V(\Sigma)$.

For this class of static configurations the field equations can be reduced to a system of three ordinary nonlinear differential equations for the metric functions $h(r)$, $f^2(r)$, and the scalar field $\phi(r)$:
\begin{subequations}  \label{eq:eqs1}
\begin{align} 
&(d-3)(\gamma-f^2)-r(h'f^2+(f^2)')=\frac{2\Lambda }{d-2}r^2,  \label{eqs1a}\\
&h'=\frac{\kappa }{(d-2)}r(\phi')^2,\label{eqs1b}\\
&\phi'=\frac{c_0}{e^hf^2r^{d-2}}\label{eqs1c}.
\end{align}
\end{subequations}
In the above equations ${}'$ denotes derivation with respect to $r$, and $c_0$ is an arbitrary constant that comes from the integration of the field equation (\ref{KG}). The first step in solving the system (\ref{eq:eqs1}) is to find an adequate variable that allows one to decouple these equations. Following \cite{Das:2001rk}, we define the new variable $a(r):=r^{d-3}e^h f^2$, in terms of which the system (\ref{eq:eqs1}) becomes a decoupled set of differential equations:
\begin{subequations}  \label{eq:eqs2}
\begin{align} 
&a^2\left[r\frac{a''}{a'}-\frac{2 \Lambda r^2-(d-3)(d-4)\gamma}{\frac{2 \Lambda}{d-2} r^2-(d-3)\gamma}\right]=\frac{\kappa c_0^2}{(d-2)}, \label{eq:eq2}\\
&h'=\frac{\kappa c_0^2}{(d-2)} \frac{1}{r a^2},\label{eqs2b}\\
&\phi'=\frac{c_0}{r a}\label{eqs2c}.
\end{align}
\end{subequations}
Note that Eq.\ (\ref{eq:eq2}) is not well defined when $\Lambda=\gamma=0$. In this case, Eq.\ (\ref{eqs1a}) implies the simple equation $a'=0$.

In Ref.\ \cite{Das:2001rk}, where only a negative cosmological constant was considered, and with a base manifold chosen to be the $(d-2)$-dimensional round sphere, the system (\ref{eq:eqs2}) was exactly solved for the case $d=3$, and the asymptotic spherically symmetric solution ($r\rightarrow \infty$), which belongs to the class of asymptotic solutions with $\gamma=1$, was also given for arbitrary $d$.

Depending on the values of $\Lambda$ and $\gamma$, four different cases can be recognized:
\begin{itemize}
\item[1.] $\Lambda=0, \gamma \neq 0$: Eq.\ (\ref{eq:eq2}) reduces to 
\begin{equation} \label{c1}
a^2\left[r\frac{a''}{a'}-(d-4)\right]=\frac{\kappa c_0^2}{(d-2)}.
\end{equation}
\item[2.] $\Lambda \neq 0, \gamma =0 $: Now Eq.\ (\ref{eq:eq2}) reduces to a very similar equation,
\begin{equation} \label{c2}
a^2\left[r\frac{a''}{a'}-(d-2)\right]=\frac{\kappa c_0^2}{(d-2)}.
\end{equation}
\item[3.] $\Lambda =0, \gamma =0$: This is the simplest case. Equation (\ref{eqs1a}) implies
 \begin{equation} \label{c3}
a'=0.
\end{equation}
\item[4.] $\Lambda \neq 0, \gamma \neq 0$: This is the general case of (\ref{eq:eq2}). 
\end{itemize}

Since Eqs.\ (\ref{c1}), (\ref{c2}), and (\ref{c3}) can be integrated, the first three cases can be completely solved. For the last one, as far as the authors know, there is no exact solution available. It is possible, however, to find the asymptotic solution for (\ref{eq:eq2}) in the general case, as we will show below.

In this article we focus mainly on the case of a negative cosmological constant and a Ricci-flat base manifold  $\Sigma$ (i.e., with $\gamma=0$). The analysis of the remaining cases, which include both known and new exact solutions is left for the interested readers.

\section{Exact general solution with a Ricci-flat base manifold} \label{exactsolution}

In this section we consider spacetimes with a Ricci-flat base manifold, i.e.\ with $\gamma=0$, in the presence of a negative cosmological constant, which is written in terms of the AdS radius $l$ as $\Lambda= -(d-1)(d-2)/(2l^2)$. We then find 
the general solution of (\ref{eq:eq2}) to be implicitly given by
\begin{equation} \label{eq:rsol}
r^{d-1}=a_0 (a-a_1)^{\frac{a_1}{a_1+a_2}}(a+a_2)^{\frac{a_2}{a_1+a_2}}.
\end{equation}
Here $a_0,a_1,a_2$ are integration constants. The constant $a_0$ can be set as $l^{d-1}$ without loss of generality, since the system  (\ref{eq:eqs2}) has a scale invariance $r\rightarrow \sigma r$ in the case $\gamma=0$. Thus $a$ becomes dimensionless. The constants $a_1$ and $a_2$ are related to the constant $c_0$ above and are defined so that they have the same sign. There is no restriction to assume that they are non-negative. Thus, the coordinate range $r\geq0$ implies the condition $a\geq a_1$ for the variable $a$. 

The resulting metric, after adjusting the irrelevant integration constant coming from (\ref{eqs2b}), written in terms of the new radial coordinate $a$, reads
\begin{equation}
\begin{split}
ds^2=&~-(a-a_1)^{\frac{(d-1)a_2-(d-3)a_1}{(d-1)(a_1+a_2)}}(a+a_2)^{\frac{(d-1)a_1-(d-3)a_2}{(d-1)(a_1+a_2)}}dt^2+\frac{l^2}{(d-1)^2}\frac{da^2}{(a-a_1)(a+a_2)}\\
&+l^2(a-a_1)^{\frac{2a_1}{(d-1)(a_1+a_2)}}(a+a_2)^{\frac{2a_2}{(d-1)(a_1+a_2)}}\gamma_{mn}dz^mdz^n.
\end{split}
\end{equation}
The solution for the scalar field is given by
\begin{equation} \label{eq:scalar_field_sol}
\phi(a)=\phi_0+\sqrt{\frac{d-2}{d-1}}\sqrt{\frac{a_1a_2}{\kappa(a_1+a_2)^2}}\ln\left(\frac{a-a_1}{a+a_2}\right),
\end{equation}
where $\phi_0$ is an arbitrary constant. Thus, the general solution contains three integration constants $\phi_0, a_1$ and $ a_2$, since $c_0$ can be written in terms of $a_1$ and $a_2$. It is usual to assume that $\phi_0=0$. However, we shall keep its value generic for now, and discuss its relation to the spacetime mass in Sec.\ \ref{sec:mass}. We remark that the second term in the right-hand side of (\ref{eq:scalar_field_sol}) could have positive or negative sign, which is obvious from the fact that only $\phi'^2$ enters in the Einstein equations. We have written it with a plus sign in front just for simplicity, where strictly it should go with a ``$\pm$'' sign.

A more simple expression for the solution can be obtained by defining the variable $x:=a+(a_2-a_1)/2$, and the constants $b:=(a_1+a_2)/2$ and $p:=(a_2-a_1)/(a_1+a_2)$. Then the solution can be written in the form
\begin{equation}
\begin{split} \label{eq:solx}
ds^2=&~-(x-b)^{\frac{1+(d-2)p}{(d-1)}}(x+b)^{\frac{1-(d-2)p}{(d-1)}}dt^2+\frac{l^2}{(d-1)^2}\frac{dx^2}{(x^2-b^2)}\\
&+l^2(x-b)^{\frac{1-p}{(d-1)}}(x+b)^{\frac{1+p}{(d-1)}}\gamma_{mn}dz^mdz^n,
\end{split}
\end{equation}
\begin{equation}\label{eq:solphix}
\phi(x)=\phi_0+\sqrt{\frac{d-2}{d-1}}\sqrt{\frac{1-p^2}{4\kappa}}\ln\left(\frac{x-b}{x+b}\right).
\end{equation}
Now, the constants $b$, $p$, and $\phi_0$ are the parameters of the family of solutions to this problem.\footnote{The particular solution with $p=0$ and $\Sigma=\mathbb{R}^{d-2}$ was found in \cite{Husain:2002jr}.} Note that $|p|\leq 1$, and $b\geq0$. The range $r >0$  implies that $x>b$.

The existence of curvature singularities can be shown through the Ricci scalar, which reads
\begin{equation} \label{eq:ricciscalar}
R=-\frac{(d-1)}{l^2}\left[d-(d-2)\frac{b^2(1-p^2)}{(x^2-b^2)}\right].
\end{equation}
Assuming that $b\neq0$ and $p^2\neq1$, i.e.\ when the scalar field is nontrivial, we see that there is a curvature singularity at $x=b$, which corresponds to $r=0$. Moreover, there are no horizons in this spacetime. This implies the existence of a naked singularity at the origin.

There are two special cases, $p^2=1$ and $b=0$, where the scalar field is trivial since it becomes a constant, $\phi=\phi_0$. The line element with $p=\pm1$ reads
\begin{equation} \label{eq:particular_case1}
ds^2=-\left[\left(\frac{r}{l}\right)^2\mp2b\Big(\frac{l}{r}\Big)^{d-3}\right]dt^2+\left[\left(\frac{r}{l}\right)^2\mp2b\Big(\frac{l}{r}\Big)^{d-3}\right]^{-1}dr^2+r^2\gamma_{mn}dz^mdz^n.
\end{equation}
In particular, the case $p=1$ gives the Schwarzschild-AdS black hole with a Ricci-flat event horizon, and a mass given by $(d-2)\kappa ^{-1}V(\Sigma)l^{d-3} b$ (see Sec.\ \ref{sec:mass}). The case $p=-1$ describes a naked singularity.

In the other special case, $b=0$, the line element reduces to
\begin{equation} \label{eq:btz0}
ds^2=-\frac{r^2}{l^2}dt^2+\frac{l^2}{r^2} dr^2+r^2\gamma_{mn}dz^mdz^n,
\end{equation}
which corresponds to a nonsingular spacetime with a vanishing mass (see Sec.\ \ref{sec:mass}). This spacetime describes a constant curvature spacetime if $\Sigma$ is a locally flat space. In particular, for $d=3$ it corresponds to the massless BTZ black hole \cite{Banados:1992wn} provided the single coordinate $z$ is an angle covering $[0, 2\pi)$.

It is interesting to note that for $d=4$, Eq.\ (\ref{eq:solx}) can be thought of as a solution for the problem with plane symmetry, with the metric depending only on the coordinate of the symmetry axis. Such a problem was considered in \cite{Vuille:2007ws}, where a particular one-parameter family of solutions was found. Our solution (with three parameters) thus generalizes the result of that reference in the case of $\Lambda<0$.  

\begin{figure}[ht]
\begin{center}
\includegraphics[width=0.65\textwidth]{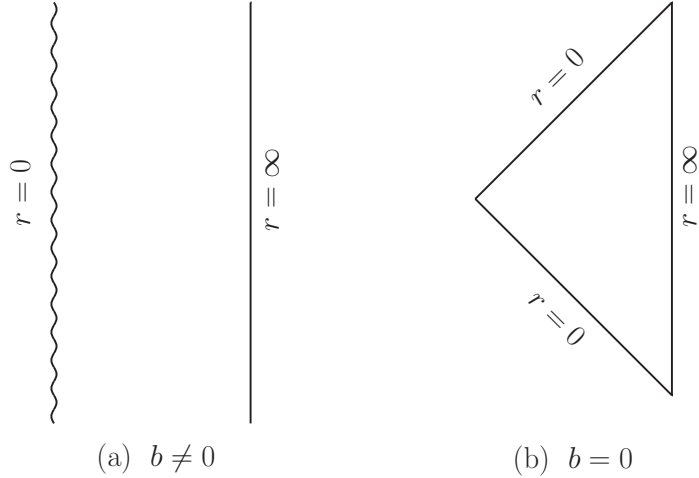}
\caption{Penrose diagrams for the spacetime (\ref{eq:solx}). The left panel (a) shows the general case $b \neq 0$. The timelike surface $r=0$ is a curvature singularity and $r= \infty$ is timelike. The right panel (b) shows the massless case $b=0$. The infinity is timelike, but the surface $r=0$ is null and does not contain a curvature singularity.}
\label{fig:penrose_diagram}
\end{center}
\end{figure}

Finally, we briefly study the causal structure of the solution (\ref{eq:solx}). The ``tortoise'' coordinate is defined by the equation
\begin{equation}
x^{*}=\frac{l}{(d-1)}\int^x \frac{dx'}{(x'-b)^{\frac{d+(d-2)p}{2(d-1)}}(x'+b)^{\frac{d-(d-2)p}{2(d-1)}}}.
\end{equation}
It can be seen that the coordinate $x^{*}$ spans a finite range of values as $x$ goes from $b$ to $\infty$, assuming that $p^2 < 1$ and $b \neq 0$. This implies that the global structure of the solution is the same as in AdS spacetime, except that the surface $r=0$ is now a curvature singularity. As in the case of AdS, future null geodesics can reach infinity, while future timelike geodesics cannot. The behavior of timelike geodesics near the singularity depends on the parameter $p$, and in fact one can show that, for $p\in (-1,-1/(d-2))$, radial timelike geodesics actually do not reach the singularity. For the case $b=0$ we have a constant scalar field and the metric reduces to (\ref{eq:btz0}). This is a massless spacetime (see Sec.\ \ref{sec:mass}) having no curvature singularity at $r=0$, and its Penrose diagram is similar to that of the BTZ massless black hole. Following \cite{HawkingEllis}, we represent the corresponding global diagrams as in Fig.\ \ref{fig:penrose_diagram}. 

\section{Asymptotic solution with a negative cosmological constant}  \label{asolution}

We now turn to study the asymptotic behavior of the solutions with a negative cosmological constant and arbitrary $\gamma$. From Eq.\ (\ref{eq:eq2}) we see that the asymptotic behavior is dominated by the cosmological constant and therefore the leading term of $a(r)$ goes like $(r/l)^{d-1}$ for $r\gg l$, which is also the leading behavior of the solution for $\gamma=0$ (see Eq.\ (\ref{eq:rsol})). Having the leading term in the asymptotic expansion for the general case, it is straightforward to find the next terms from Eq.\ (\ref{eq:eq2}). We find
\begin{equation} \label{eq:asympt}
a(r)=\left(\frac{r}{l}\right)^{d-1}+\gamma\left(\frac{r}{l}\right)^{d-3}-\mu+O\left(\Big(\frac{l}{r}\Big)^{d-1}\right),
\end{equation}
where $\mu$ is an arbitrary constant. In terms of the integration constants of the exact solution (\ref{eq:solx}) for the case $\gamma=0$, the constant $\mu$ is equal to $2bp$. With the result of Eq.\ (\ref{eq:asympt}) we obtain the asymptotic expansion of the metric for the general case:
\begin{equation}
\begin{split} \label{eq:asympmetric}
ds^2=&~-\left[\left(\frac{r}{l}\right)^2+\gamma-\mu\Big(\frac{l}{r}\Big)^{d-3}+O\left(\Big(\frac{l}{r}\Big)^{2(d-2)}\right)\right]dt^2\\
&+\left[\left(\frac{r}{l}\right)^2+\gamma-\mu\Big(\frac{l}{r}\Big)^{d-3}+O\left(\Big(\frac{l}{r}\Big)^{2(d-2)}\right)\right]^{-1}dr^2+r^2\gamma_{mn}dz^mdz^n.
\end{split}
\end{equation}
Note that the functions in square brackets in the above equation are equal only up to order $(l/r)^{d-3}$, but differ to higher orders. From (\ref{eq:asympmetric}), and from the fact that $R\to -d(d-1)/l^2$ as $r\to\infty$, we see that the spacetime appears to be asymptotically AdS. Strictly speaking, it is asymptotically AdS only for $\gamma=1$ and $\Sigma$ fixed as the $(d-2)$-sphere. For $\gamma=0,-1$, the spacetimes are asymptotically \textit{locally} AdS spacetimes only if the base manifold $\Sigma$ is a constant curvature space. This condition is automatically satisfied in four and five spacetime dimensions. In these dimensions $\Sigma$ is locally isomorphic to the sphere $\mathbb{S}^{d-2}$, the hyperbolic manifold $\mathbb{H}^{d-2}$, or the Euclidean space $\mathbb{R}^{d-2}$, for $\gamma=1,-1,0$, respectively.

Replacing (\ref{eq:asympt}) in (\ref{eqs2c}), we compute the asymptotic form of the scalar field,
\begin{equation} \label{eq:asympphi}
\phi=\phi_0-\phi_1\Big(\frac{l}{r}\Big)^{d-1}+O\left(\Big(\frac{l}{r}\Big)^{d+1}\right),
\end{equation}
where $\phi_0$ and $\phi_1$ are arbitrary constants. For the exact solution (\ref{eq:solphix}), $\phi_1=\sqrt{\frac{d-2}{d-1}}\sqrt{\frac{1-p^2}{\kappa}}b$. The full family of asymptotic solutions is thus parametrized by the three constants $\mu$, $\phi_0$, and $\phi_1$.

\section{Mass} \label{sec:mass}

Starting from the asymptotic behavior of the metric and the scalar field, we now turn to the problem of computing the mass of these configurations. We address this issue following the Regge-Teitelboim approach\footnote{Using different methods, finite values for the mass were computed in \cite{Das:2001rk} for the cases of $d=3,4,5$.} \cite{Regge:1974zd}. In general, for the action considered here, the variation of the conserved charges corresponding to the asymptotic symmetries defined by the vector $\xi=(\xi^t,\xi^i)$, is given by
\begin{equation}
\delta Q(\xi)=\delta Q_G(\xi)+\delta Q_{\phi}(\xi),
\end{equation}
where \cite{Henneaux:2006hk}
\begin{equation} \label{eq:Q_G}
\delta Q_G(\xi)=\frac{1}{2\kappa}\int d^{d-2}S_l G^{ijkl}(\xi^{\bot}\delta g_{ij;k}-\xi^{\bot}_{~,k}\delta g_{ij})+\int d^{d-2}S_l(2\xi_k\delta\pi^{kl}+(2\xi^k\pi^{jl}-\xi^l\pi^{jk})\delta g_{jk}),
\end{equation}
\begin{equation} \label{eq:Q_phi}
\delta Q_{\phi}(\xi)=-\int d^{d-2}S_l(\xi^{\bot}g^{1/2}g^{lj}\partial_j\phi\delta\phi+\xi^l\pi_{\phi}\delta\phi),
\end{equation}
are, respectively, the gravitational and scalar field contributions. Here $g_{ij}$ denotes the components of the $(d-1)$-spatial metric, $\pi^{ij}$ are their conjugate momenta, and $\pi_{\phi}$ is the momentum associated with $\phi$. We have also defined $\xi^{\bot}=\xi^t\sqrt{-g_{tt}}$, and
\begin{equation}
G^{ijkl}\equiv \frac{1}{2}g^{1/2}(g^{ik}g^{jl}+g^{il}g^{jk}-2g^{ij}g^{kl}).
\end{equation}
In the static case all the momenta vanish, and the relevant asymptotic symmetry corresponds to the vector $\partial_t$. The mass is the conserved charge associated with this symmetry. We then write the variation of the mass as $\delta M=\delta Q(\partial_t)=\delta M_G+\delta M_{\phi}$, and from Eqs.\ (\ref{eq:asympmetric}) and (\ref{eq:asympphi}) we obtain
\begin{equation} \label{eq:delta_mg}
\delta M_G=-\lim_{r\to\infty}\frac{(d-2)}{2\kappa}V(\Sigma)\frac{r^{d-2}}{l}(g^{rr})^{-1/2}\delta g^{rr}=\frac{(d-2)}{2\kappa}V(\Sigma)l^{d-3}\delta\mu,
\end{equation}
\begin{equation} \label{eq:delta_mphi}
\delta M_{\phi}=-\lim_{r\to\infty} V(\Sigma)\frac{r^{d-1}}{l}(g^{rr})^{1/2}\phi'\delta\phi=(d-1)V(\Sigma)l^{d-3}\phi_1\delta\phi_0,
\end{equation}
where $V(\Sigma)$ denotes the volume of the Einstein base manifold.

The next step is to integrate Eqs.\ (\ref{eq:delta_mg}) and (\ref{eq:delta_mphi}) in order to obtain the value of the mass $M$. The gravitational contribution can be directly integrated, giving the result
\begin{equation} \label{eq:mass1}
M_G=\frac{(d-2)}{2\kappa}V(\Sigma)l^{d-3}\mu.
\end{equation}
The above expression corresponds to the standard mass formula for a spacetime with a metric of the form (\ref{eq:asympmetric}) in vacuum. 

The issue of the scalar field contribution to the mass is more subtle, since now $\delta M_{\phi}$ depends on the two integration constants $\phi_0$ and $\phi_1$ through the combination $\phi_1\delta\phi_0$. In general, the integration of this variation requires a functional relation between $\phi_0$ and $\phi_1$. Consequently, the scalar field contribution $M_{\phi}$ will be determined by this relation, and so will be the total mass as well. Indeed, the same situation occurs in the case of a \textit{massive} scalar field on an asymptotically AdS spacetime. In this setup, the Klein-Gordon equation leads to
\begin{equation} \label{eq:phimass}
\phi \sim\frac{\phi_0}{r^{\Delta_-}}-\frac{\phi_1}{r^{\Delta_+}},
\end{equation}
for large $r$, where
\begin{equation}
\Delta_{\pm}= \frac{(d-1)}{2}\left(1\pm\sqrt{1+\frac{4 l^2 m^2}{(d-1)^2}}\right),
\end{equation}
with $m$ being the scalar field mass. For a massless scalar field we have $\Delta_-=0$ and $\Delta_+=d-1$. In the most general case, $\phi_0$ and $\phi_1$ are functions depending on time and also on the $d-2$ coordinates of the base manifold. For the analysis developed here, it is enough to consider $\phi_0$ and $\phi_1$ as integration constants. In Ref.\ \cite{Henneaux:2006hk} it was shown that the term 
\begin{equation}
\Delta_- \phi_0 \delta \phi_1+\Delta_+ \phi_1 \delta \phi_0
\end{equation}
gives a contribution to the mass, and can be integrated assuming a functional relation between $\phi_0$ and $\phi_1$. The functional relation is fixed as
\begin{equation} \label{relation}
\phi_0= \alpha \phi_1^{\frac{\Delta_-}{\Delta_+}}, 
\end{equation}
with $\alpha$ an arbitrary constant with no variation, after one imposes that this functional relation is preserved by the asymptotic AdS symmetry. In fact, the $\xi^r$ component of the vector $\xi$ defining the AdS asymptotic group, which is linear in $r$, fixes the relation (\ref{relation}). This symmetry corresponds to an asymptotic scaling invariance, as we discuss next.

Here we are considering solutions having only a subset of the AdS symmetries at infinity. In particular, system (\ref{eq:eqs1}) possesses a scaling symmetry, provided that $\gamma=0$, given by 
\begin{equation} \label{eq:scale_transf}
\begin{split}
\tilde{r}=&~\sigma r,\\
\tilde{f}(\tilde{r})=&~\sigma f(r),\\
e^{\tilde{h}(\tilde{r})}=&~\sigma^{-(d-1)}e^{h(r)},\\
\tilde{\phi}(\tilde{r})=&~\phi(r),
\end{split}
\end{equation}
where $\sigma$ is a positive constant.\footnote{This can be also extended to rotating and charged configurations in $d=3$ \cite{Banados:2005hm}.} This symmetry is also an asymptotic symmetry even when $\gamma \neq 0$. For an infinitesimal scaling $\sigma=1+\varepsilon$ one finds 
\begin{equation}
\delta_{\sigma}\phi=-r \varepsilon \phi' = -\varepsilon (d-1) \phi_1\Big(\frac{l}{r}\Big)^{d-1}+O\left(\Big(\frac{l}{r}\Big)^{d+1}\right).
\end{equation}
On the other hand, the functional variation at infinity goes as 
\begin{equation} \label{eq:varasympphi}
\delta\phi=\delta\phi_0-\delta\phi_1\Big(\frac{l}{r}\Big)^{d-1}+O\left(\Big(\frac{l}{r}\Big)^{d+1}\right),
\end{equation}
as follows from (\ref{eq:asympphi}). Thus, if one restricts the functional variations at infinity to be compatible with the variations generated by an infinitesimal scaling, the condition
\begin{equation} \label{condition}
\delta\phi_0=0
\end{equation}
must hold. This condition is satisfied provided that $\phi_0$ is a constant without variation. The same conclusion can be obtained from (\ref{relation}) with $\Delta_{-}=0$. In this way, under the condition (\ref{condition}) we have $\delta M_{\phi}=0$, and then the total mass is just $M=M_G$. In summary, if the functional variations at infinity are restricted to be compatible with those coming from the scaling invariance, we can conclude that the scalar field does not contribute to the mass. On the contrary, for a generic variation of the scalar field at infinity one should expect a nonzero contribution. We remark that the condition $\delta\phi_0=0$ amounts to fixing Dirichlet boundary conditions, and it is interesting to note that, even with a different choice of boundary conditions, the mass of the spacetime will still be finite.

The above results allow us to calculate the mass of the exact solution (\ref{eq:solx}-\ref{eq:solphix}). In this case we have $\mu=2 b p$ and considering $\phi_0$ fixed (i.e.\ $\delta \phi_0=0$), Eq.\ (\ref{eq:mass1}) gives
\begin{equation}
M= M_G=\frac{(d-2)}{\kappa}V(\Sigma)l^{d-3} b p.
\end{equation}

\section{Final remarks} \label{sec:remarks}

In this paper we have studied static configurations in Einstein gravity minimally coupled to a massless scalar field, focusing our attention mainly on solutions with a Ricci-flat base manifold in the presence of a negative cosmological constant. However, the case of a positive cosmological constant, in arbitrary dimensions and with $\gamma=0$, can be treated using the same method we used for the decoupling of the system of equations (\ref{eq:eqs1}). These exact solutions with $\Lambda >0$ may have a cosmological interpretation as occurs in the $\Lambda=0$ four-dimensional case \cite{Abdolrahimi:2009dc}, where the existence of the so-called ``Fisher universes'' can be established.

The power of having a procedure for decoupling the system of differential equations can be clearly seen in the case of $\Lambda=0=\gamma$. The equation for the variable $a$ is just $a'=0$, yielding the line element
\begin{equation}
ds^2=-\left(\frac{r}{r_0}\right)^{q-(d-3)}dt^2+\left(\frac{r}{r_0}\right)^{q+(d-3)}dr^2+r^2\gamma_{mn}dz^mdz^n,
\end{equation}
and the scalar field
\begin{equation}\label{eq:solphix2}
\phi(r)=\phi_0 \pm\sqrt{\frac{(d-2)q}{\kappa}}\ln\left(\frac{r}{r_0}\right),
\end{equation}
where $r_0$, $q\geq 0$, and $\phi_0$ are integration constants. This solution reduces in three dimensions to the result found in \cite{Virbhadra:1994xz}, and it contains a naked singularity located at $r=0$.

The previous simple case reveals again a relevant aspect of the solutions we present, namely the presence of a naked singularity at the origin. This is a well known result for the aymptotically flat case \cite{Janis:1968zz}, and it seems to be a generic property of scalar field spacetimes. Thus, in the light of the cosmic censorship conjecture this class of solutions would be ruled out on physical grounds. The validity of this conjecture, however, is an open problem \cite{Penrose:1999vj}, and a number of counterexamples exist in which plausible models of gravitational collapse lead to a naked singularity (see e.g.\ \cite{Harada:2001nj}). 

The mass of the asymptotically locally anti-de Sitter solutions (as well as those with a base manifold with nonconstant curvature) has, in general, a nontrivial contribution coming from the scalar field $M_{\phi}$. The precise form of this contribution depends on the relation between the leading terms of the asymptotic expansion of the scalar field. However, if one restricts the variations at infinity to those preserving the scaling invariance [Eqs.\ (\ref{eq:scale_transf})], then one finds that $M_{\phi}=0$, since compatibility with the scaling symmetry forces the integration constant $\phi_0$ to be a fixed parameter (Dirichlet boundary conditions) of the family of solutions considered.  

Whatever the case may be, the fact that these spacetimes containing naked singularities always have finite energy may imply that they are physically acceptable \cite{Das:2001rk}. This interesting property also gives rise to the possibility of a semiclassical phase transition from a black hole to a naked singularity. In fact, it was shown that this transition occurs in three spacetime dimensions \cite{Das:2001wu}, and it would be of interest to study the existence of such transition in higher dimensions.

\acknowledgments The authors thank Hideki Maeda and Ricardo Troncoso for useful discussions. This work has been partially funded by the Fondecyt grants 1121031, 1085322, 1095098, 1100755 and by the Conicyt grant ACT-91: ``Southern Theoretical Physics Laboratory'' (STPLab). The Centro de Estudios Cient\'{\i}ficos (CECs) is funded by the Chilean Government through the Centers of Excellence Base Financing Program of Conicyt.


\end{document}